\documentclass[prb,onecolumn,showpacs,amssymb,superscriptaddress,notitlepage,nofootinbib]{revtex4-2}

\usepackage{graphicx}
\usepackage{amsmath}
\usepackage{amsfonts}
\usepackage{amsthm}
\usepackage{amssymb}
\usepackage{amsbsy}
\usepackage{wasysym}
\usepackage{bm}
\usepackage{mathrsfs}
\usepackage{color}
\usepackage{hyperref}
\usepackage{braket}
\usepackage{verbatim}
\setcitestyle{super}

\newcommand*{\citen}[1]{%
 \begingroup
    \romannumeral-`\x 
    \setcitestyle{numbers}%
    \cite{#1}%
 \endgroup   
}

\date{\today}
\begin{document}

\title{Quasiparticle kinetic theory for Calogero models}

\author{Vir B. Bulchandani}
\affiliation{Princeton Center for Theoretical Science, Princeton University, Princeton, New Jersey 08544, USA}
\author{Manas Kulkarni}
\affiliation{International Centre for Theoretical Sciences, Tata Institute of Fundamental Research, Bengaluru  560089, India}
\author{Joel E. Moore}
\affiliation{Department of Physics, University of California, Berkeley, California 94720 USA}
\affiliation{Materials Sciences Division, Lawrence Berkeley National Laboratory, Berkeley, California 94720, USA}
\author{Xiangyu Cao}
\affiliation{Laboratoire de Physique de l'Ecole Normale Sup\'erieure, ENS, Universit\'e PSL,
CNRS, Sorbonne Universit\'e, Universit\'e de Paris, 75005 Paris, France} 

\begin{abstract}
We show that the quasiparticle kinetic theory for quantum and classical Calogero models reduces to the free-streaming Boltzmann equation. We reconcile this simple emergent behaviour with the strongly interacting character of the model by developing a Bethe-Lax correspondence in the classical case. This demonstrates explicitly that the freely propagating degrees of freedom are not bare particles, but rather quasiparticles corresponding to eigenvectors of the Lax matrix. We apply the resulting kinetic theory to classical Calogero particles in external trapping potentials and find excellent agreement with numerical simulations in all cases, both for harmonic traps that preserve integrability and exhibit perfect revivals, and for anharmonic traps that break microscopic integrability. Our framework also yields a simple description of multi-soliton solutions in a harmonic trap, with solitons corresponding to sharp peaks in the quasiparticle density. Extensions to quantum systems of Calogero particles are discussed.
\end{abstract}

\maketitle
\section{Introduction}

The emergence of macroscopic fluid behaviour from the chaotic motion of microscopic particles is a truth universally acknowledged, whose mathematical derivation nevertheless continues to present insurmountable difficulties. Aside from certain tractable classical cases~\cite{spohn2012large} and a handful of quantum examples~\cite{erdHos2010derivation}, microscopic derivations of hydrodynamics remain beyond reach at present. A significant conceptual advance in this area was the realization that large-scale dynamics in a wide range of extended integrable systems~\cite{Castro-Alvaredo2016,Bertini2016} could be reduced to a kinetic theory of solitons~\cite{Percus69,Zakharov71,Boldrighini83,El2005,DoyonSoliton,BVKM1,GopDiff,GopSuperdiff}. This synthesis has created a powerful and increasingly rigorous~\cite{Panfil,Fagotti,CollRate,Poszgay} tool for understanding the emergence of hydrodynamic behaviour in realistic models from microscopic first principles, that moreover yields excellent agreement with state-of-the-art experimental results~\cite{Dubail,Scheie2021,malvania2020generalized}.

One experimentally important question that remains unclear, despite several recent related studies~\cite{GHDnote,CBMTrap,CauxQNC,BastianelloTrap,bastianello2021hydrodynamics,durnin2021diffusive}, is the extent to which a kinetic theory of solitons continues to provide an accurate description in the presence of integrability-breaking trapping potentials. From this viewpoint, the family of Calogero-type models, which are integrable one-dimensional systems that remain integrable in the presence of certain carefully chosen trapping potentials~\cite{kulkarni2017emergence}, are natural objects of study. Moreover, their (zero temperature) hydrodynamics has been explored in some depth~\cite{OldCalog1,OldCalog2,stone2008classical,OldCalog3}.

In this paper, we develop a quasiparticle kinetic theory for Calogero models. We first derive the kinetic theory of the quantum Calogero model using established thermodynamic Bethe ansatz techniques~\cite{Bertini2016,Castro-Alvaredo2016}, showing that it reduces to a free-streaming Boltzmann equation. We then obtain the kinetic theory of the classical Calogero model as a semiclassical limit of the kinetic theory of the quantum model, following the analogous procedure for the Toda lattice~\cite{BCM,Theo,GrunerBauer1988} (see also Refs. ~\citen{Bastianello2018,Spohn_2019,DoyonToda,CBS,spohn2021hydrodynamic}), which also turns out to be of non-interacting Boltzmann form. For the classical Calogero model, we explain this simple behaviour using the Bethe-Lax correspondence proposed in previous work~\cite{BCM}. This provides an independent check on the kinetic theory description and illustrates that the freely streaming degrees of freedom are not bare particles, as one might expect from the absence of velocity dressing, but instead quasiparticles that can be identified with eigenvectors of the Lax matrix.

We then study the validity of the quasiparticle kinetic theory, as augmented by a na{\"i}ve Boltzmann force term, in the presence of external trapping potentials. For the integrability-preserving case of a harmonic trapping potential (also known as the Calogero-Moser model) we find that this kinetic theory captures the finite-temperature dynamics to within numerical accuracy, including non-trivial features such as perfect revivals and soliton excitations~\cite{OldCalog3}. For integrability-breaking anharmonic potentials, we again find excellent agreement with numerics. This is attributed to an unusual robustness of integrability of the Calogero dynamics to external trapping potentials. We close with some remarks on the dynamics of trapped quantum Calogero particles.

\section{Kinetic theory}
\subsection{Quantum Calogero model}\label{sec:quantum}
Consider the $N$-particle, quantum (rational) Calogero model, with Hamiltonian
\begin{equation}
H = \sum_{i=1}^N -\frac{1}{2} \partial_i^2 + \sum_{i<j} \frac{g}{(x_i-x_j)^2} \,,
\end{equation}
where we set $\hbar = m = 1$ and $g = \alpha(\alpha-1)$ for some $\alpha > 0$. This model is integrable, and its exact spectrum can be obtained from Sutherland's method of asymptotic Bethe ansatz\cite{Sutherland}. The two-particle phase shift is given by
\begin{equation}
\varphi(k) = \pi (\alpha-1) \mathrm{sgn}(k) \,,
\end{equation}
implying that the differential phase shift is
\begin{equation}
\label{diff}
K_{k,k'} = \frac{1}{2\pi}\varphi'(k-k') = (\alpha-1)\delta(k-k') \,.
\end{equation}
For states in thermal equilibrium with chemical potential $\mu$ and inverse temperature $\beta$, it follows by thermodynamic Bethe ansatz\cite{YY} that quasiparticle energies satisfy the Yang-Yang equation
\begin{align}
\label{YY}
\epsilon_k = \frac{k^2}{2}-\mu + \frac{1}{\beta}\int_{-\infty}^{\infty} dk' \, K_{k,k'} \ln{(1+e^{-\beta \epsilon_{k'}})} \,,
\end{align}
while the total density of states satisfies
\begin{equation}
\label{DoS}
\rho_k^t + \int_{-\infty}^{\infty} dk' \, K_{k,k'} \theta_{k'} \rho_{k'}^t = \frac{1}{2\pi} \,,
\end{equation}
with Fermi factors given by $\theta_k = (1+e^{\beta \epsilon_k})^{-1}$. For the phase shift Eq. \eqref{diff}, different values of pseudomomentum $k$ decouple. The Yang-Yang equation can thus be expressed as a transcendental equation for $\theta_k$, that fixes the occupied density of states $\rho_k = \rho^t_k \theta_k$:
\begin{align}
\frac{\theta_k}{\theta_k + (1-\theta_k)^\alpha} &= \frac{1}{1+e^{\beta(k^2/2-\mu)}} \,, \\
\label{DoS2}
\rho_k &= \frac{1}{2\pi} \frac{\theta_k}{1+(\alpha-1)\theta_k} \,.
\end{align}
For $\alpha =0$, Eq. \eqref{DoS2} recovers the occupation numbers for free bosons, while for $\alpha=1$, it yields the occupation numbers for free fermions. For other values of $\alpha$, these equations describe the thermodynamics of free anyons\cite{Isakov}.

To proceed from thermodynamic Bethe ansatz to the generalized hydrodynamics of the Calogero model, we require an expression for the quasiparticle group velocity on a given equilibrium state\cite{Castro-Alvaredo2016,Bertini2016}. (Note that for the Calogero model, the kinetic theory description is expected to be equivalent to a generalized hydrodynamics consisting of countably many hydrodynamic equations, with no subtleties arising due to vacuum modes~\cite{superdiffusion}.) Recall that the derivatives of energy and momentum for a quasiparticle excitation on a given equilibrium state with Fermi factors $\{\theta_k\}_{k\in\mathbb{R}}$ satisfy
\begin{eqnarray}
\nonumber \epsilon'_k +\int_{-\infty}^{\infty} dk' \, K_{k,k'}\theta_{k'}\epsilon'_{k'} &= k \,, \\
{p^{\mathrm{dr}}}'_k +\int_{-\infty}^{\infty} dk' \, K_{k,k'}\theta_{k'}{p^{\mathrm{dr}}}'_{k'} &= 1 \, .
\end{eqnarray}
For the differential phase shift Eq. \eqref{diff}, these reduce to
\begin{align}
(1+ (\alpha-1)\theta_k)\epsilon'_k &= k \,, \\
(1+ (\alpha-1)\theta_k){p^{\mathrm{dr}}}'_k &= 1 \,.
\end{align}
It follows that the quasiparticle group velocity in a given equilibrium state simply equals the bare velocity:
\begin{equation}
\label{vel}
v_k = \epsilon'_k/{p^{\mathrm{dr}}}'_k = k \,.
\end{equation}
Thus the Bethe-Boltzmann equation for the Calogero model reduces to the free-streaming Boltzmann equation,
\begin{equation}
\label{BBQCalogero model}
\partial_t \rho_k + k \partial_x \rho_k = 0 \,.
\end{equation}
Note that state independence of the group velocity Eq. \eqref{vel} additionally implies that its fluctuations vanish on equilibrium states, and hence that the Navier-Stokes correction to Eq. \eqref{BBQCalogero model} vanishes\cite{DeNardis2018,Gopalakrishnan2018}. This implies that the Calogero model is non-interacting in the sense of Spohn~\cite{SpohnIntNonInt}, despite its singular inter-particle interactions.

\subsection{Classical Calogero model}\label{sec:classical}
We now derive the thermodynamic Bethe ansatz and kinetic theory for the classical Calogero model by passing to the semiclassical limit, adapting the analogous derivation for the Toda lattice\cite{Theo,GrunerBauer1988,BCM}. Our starting point is the quantum Calogero model with Planck's constant restored, namely
\begin{equation}
H =  \sum_{i=1}^N -\frac{1}{2} (\hbar\partial_i)^2 + \sum_{i<j} \frac{\hbar^2 \alpha(\alpha-1)}{(x_i-x_j)^2} \,.
\end{equation}
The semiclassical limit is obtained by expressing the quantum TBA equations in terms of classical momentum $p = \hbar k$, before taking $\hbar \to 0$ with $\ell = \hbar \alpha$ constant~\cite{Choquard}. To obtain a finite free energy in this limit, we define classical energies and chemical potentials
\begin{equation}
\epsilon_p^{cl} = \epsilon_p + \frac{1}{\beta}\ln{\hbar/\ell}, \quad \mu^{cl} = \mu - \frac{1}{\beta}\ln{\hbar/\ell} \,.
\end{equation}
For non-zero $\hbar$, the Yang-Yang equation and dressing equations can be written exactly in terms of $p$, as\cite{BCM}
\begin{align}
\epsilon^{cl}_p = \frac{p^2}{2} - \mu^{cl} + \frac{1}{\beta \hbar}\int_{-\infty}^{\infty} dp' \, K_{p/\hbar,p'/\hbar} \ln{(1+\hbar\ell^{-1} e^{-\beta\epsilon^{cl}_{p'}})}
\end{align}
and
\begin{align}
{\epsilon^{cl}}'_p +\int_{-\infty}^\infty dp' \, K_{p/\hbar,p'/\hbar}(\theta_{p'}/\hbar){\epsilon^{cl}}'_{p'} &= p \,, \\
{p^{\mathrm{dr}}}'_p +\int_{-\infty}^\infty dp' K_{p/\hbar,p'/\hbar}(\theta_{p'}/\hbar){p^{\mathrm{dr}}}'_{p'} &= 1 \,.
\end{align}
The classical total density of states satisfies $2 \pi \hbar \widetilde{\rho}^t_p = {p^{\mathrm{dr}}}'_p$. In the limit $\hbar \to 0$, the phase shift is given by
\begin{equation}
K^{cl}_{p,p'} = \lim_{\hbar \to 0} K_{p/\hbar,p'/\hbar} = \lim_{\hbar \to 0} \hbar(\alpha-1) \delta(p-p') = \ell \delta(p-p') \,,
\end{equation}
and the Yang-Yang equation becomes
\begin{equation} \label{eq:YY}
\epsilon^{cl}_p = \frac{p^2}{2m} - \mu^{cl} + \frac{1}{\beta}e^{-\beta \epsilon^{cl}_p} \,.
\end{equation}
Note that this is independent of the interaction strength, $\ell$. In fact, so are the dressing equations, and read
\begin{align}
\label{classdress1}
(1+e^{-\beta \epsilon^{cl}_p}) {\epsilon^{cl}}'_p  &= p \,, \\
\label{classdress2}
(1+e^{-\beta \epsilon^{cl}_p}) {p^{\mathrm{dr}}}'_p  &= 1 \,.
\end{align}
This reflects the fact that the leading asymptotic kinematics in the classical Calogero model~\cite{Calogero} is independent of $\ell$. However, the classical density of occupied states does depend on $\ell$, and is given by
\begin{equation}
\widetilde{\rho}_p = \lim_{\hbar \to 0} \widetilde{\rho}_p^t \theta_p = \frac{1}{2\pi} \frac{1}{1+e^{\beta \epsilon^{cl}_p}}\frac{1}{\ell} \,. \label{eq:rho_p_classical}
\end{equation}
Finally, we note that the classical dressing equations Eqs. \eqref{classdress1} and \eqref{classdress2} imply an effective velocity
\begin{equation}
v_p = {\epsilon^{cl}}'_p/{p^{\mathrm{dr}}}'_p = p
\end{equation}
that is free-particle like. The resulting kinetic theory takes the form of a free-streaming Boltzmann equation,
\begin{equation}
\label{eq:class_boltzmann}
\partial_t \widetilde{\rho}_p + p \partial_x \widetilde{\rho}_p = 0 \,,
\end{equation}
as in the quantum case.

In the presence of an external trapping potential, the na{\"i}ve modification of Eq. \eqref{eq:class_boltzmann} by a Boltzmann force term reads
\begin{equation}
\label{eq:minmod}
\partial_t \widetilde{\rho}_p + p \partial_x \widetilde{\rho}_p - V'(x) \partial_p \widetilde{\rho}_p = 0 \,.
\end{equation}
The regime of validity of this equation is not immediately clear, as the presence of an external trapping potential is expected in general to break integrability and generate finite quasiparticle lifetimes, ultimately invalidating any description based on a kinetic theory of quasiparticles.  This crossover time-scale for the onset of chaos, which depends on the initial condition, was previously obtained for the classical hard rod model with integrability broken by a harmonic trap~\cite{CBMTrap}.

However, Calogero-type models are unusual among integrable models because they remain integrable in the presence of certain carefully chosen external trapping potentials. For the rational Calogero model considered here, the known integrability-preserving potentials take the form~\cite{PolychronakosReview}
\begin{align}
\label{eq:arbtrap}
V(x) = ax^4 + bx^3 + cx^2 + d, \quad a,b,c,d \in \mathbb{R} \,.
\end{align}
For this restricted class of potentials, the validity of Eq. \eqref{eq:minmod} is more plausible. This result also suggests a qualitative robustness of Calogero particles to integrability-breaking by generic smooth potentials $V(x)$, since approximating $V(x)$ by the first four terms of a Taylor expansion about any given point $x=x_0$ will always yield an integrable model.

\section{Bethe-Lax correspondence}\label{sec:bethelax}
The classical Calogero model exhibits a Lax pair formalism\cite{PolychronakosReview}, just as for the classical Toda model\cite{Flaschka}. In previous work, it was noted that conserved quasiparticles in the Toda model could be understood as eigenvectors of the Lax matrix\cite{BCM, CBS}. In particular, it was found that the equation of state for quasiparticle currents\cite{DoyonToda,BCM} could be interpreted as a statement about thermal averages of eigenvectors of the Lax matrix. It is natural to conjecture that a similar correspondence holds for the classical Calogero model. We shall show that this is indeed the case, and moreover, that the Bethe-Lax correspondence explains why the current density of states takes the simple, non-interacting form
\begin{equation}
\widetilde{\rho}^J_p = p\widetilde{\rho}_p \,. \label{eq:Jandp}
\end{equation}

\subsection{From matrix models to Lax pairs: a reminder}
It will be useful to recall that the classical Calogero model can be obtained from the Hamiltonian reduction of a Hermitian matrix model\cite{PolychronakosReview}, defined by the Lagrangian
\begin{equation}
    \mathcal{L} = \frac12 \mathrm{Tr}(\dot{M} \dot{M}^\dagger) - \mathrm{Tr} V(M)\,,
\end{equation}
where $M$ is an $N\times N$ Hermitian matrix 
and $V(M)$ is usually a polynomial function (the potential). The equations of motion read 
\begin{equation}
\label{eq:matrixEOM}
    \frac{dM}{dt} = \Lambda \,,\,
    \frac{d \Lambda} {dt} = F (M) \,,
\end{equation}
where $F(x) = -V'(x)$ is the force. 

The $SU(N)$ symmetry of the action means that one should be able to describe the dynamics solely in terms of the eigenvalues of $M$. To this end, let $U$ denote the dynamical unitary transformation diagonalizing $M$:
\begin{equation}
    M = U^\dagger X U \,, X = \mathrm{diag} (x_1,\dots, x_N ) \,.
\end{equation}
Then the motion of $M$ has a ``radial'' and ``angular'' part
\begin{equation}
    \frac{d}{dt}M = U^\dagger \left( \dot{X} + i [X, A] \right)  U  \,,\, i A := \dot{U}U^{\dagger} \,. \label{eq:radial_angular}
\end{equation}
where $A$ is the Hermitian matrix generating the angular motion. Letting $L$ denote the rotated version of $\Lambda$,
\begin{eqnarray}
\label{eq:defL}
\Lambda = U^\dagger L U  \,,\, \frac{d}{dt}\Lambda = U^\dagger \left( \dot{L} + i [L, A] \right)  U
\end{eqnarray}
the equations of motion for $M$ and $\Lambda$ can be rewritten in terms of $X$ and $L$ as
\begin{align}
\label{eq:redEOM1}
\dot{X} + i[X,A] &= L \,, \\
\label{eq:redEOM2}
\dot{L} + i[L,A] &= F(X)
\end{align}
(note that we used $U F(M) U^{\dagger} = F(X)$, which assumes that $F(M)$ is analytic). To see how the Calogero model emerges explicitly, we need to fix the ``angular momentum'' associated with the $SU(N)$ symmetry. The conserved current is 
\begin{equation}
    J = -i [M, \dot{M} ] \Rightarrow \dot{J} = 0 \,.
\end{equation}
The Calogero model comes from choosing a traceless matrix with one-dimensional kernel for $J$:
\begin{equation}
    J = \ell (v v^\dagger - I) \,,\,  v^\dagger v = N \,.
\end{equation}
Then, since
\begin{equation}
    J = -i [M, \dot{M}] = U^\dagger [X, [X, A]] U  \,,
\end{equation}
we have
\begin{equation}
     [X, [X, A]] = U J U^{\dagger} = \ell (u u^\dagger - I) \,, 
     u := Uv \,.
\end{equation}
Now the LHS has zero diagonal elements, so $|u_i|^2 = 1$, and we can set $u_i = 1$. Then the above equation implies $A_{ij} = \ell x_{ij}^{-2}$ for $i \neq j$ ($x_{ij} := x_i - x_j$). The diagonal elements are set by $u^\dagger A = 0$, so that we have
\begin{equation}
\label{eq:defA}
    A_{ij} = -\delta_{ij} \ell \sum_{k \neq i} x_{ik}^{-2} + (1-\delta_{ij}) \ell x_{ij}^{-2} \,.
\end{equation}
It follows from \eqref{eq:redEOM1} that 
\begin{equation}
\label{eq:defL2}
    L_{ij} =  \delta_{ij} \dot{x}_i + (1-\delta_{ij}) i\ell x_{ij}^{-1} \,.
\end{equation}
We have thus recovered the Lax pair for the classical Calogero model. 

\subsection{Statement of the correspondence}
In the absence of an external potential, the force $F(X)$ vanishes and the flow of the Lax matrix is isospectral, i.e. its eigenvalues are conserved. The question then arises of how these conserved eigenvalues relate to the conserved quasiparticle momenta in the classical Boltzmann description, Eq. \eqref{eq:class_boltzmann}. As in previous work on the Toda lattice~\cite{BCM}, we argue that these quantities can be identified, and moreover that the \emph{eigenvectors} of the Lax matrix are naturally interpreted as quasiparticles, giving rise to a ``Bethe-Lax correspondence'' between the semiclassical Bethe ansatz and the spectral data of the classical Lax matrix.

Concretely, note that if $V(M) = 0$ in Eq. \eqref{eq:matrixEOM}, then $\dot{M} = \Lambda$ is conserved and by $SU(N)$ symmetry can be chosen to be diagonal:
\begin{equation}
    \Lambda = \mathrm{diag}(\lambda_1, \dots, \lambda_N) \,,\, 
    \dot{\Lambda} = 0 \,.
\end{equation}
Since $L$ and $\Lambda$ are related by the similarity transformation Eq. ~\eqref{eq:defL}, it follows that $\{\lambda_i\}_{i=1}^N$ are the eigenvalues of $L$. Also, since $L = U \Lambda U^\dagger$, we can directly identify the columns of $U$ with the eigenvectors of $L$, yielding the expressions
\begin{equation}
    U = \sum_j | \lambda_j  \rangle \langle  j | \,,\, 
    L  | \lambda_j  \rangle = \lambda_j   | \lambda_j  \rangle \,,\, \frac{d}{d t}
    | \lambda_j  \rangle = i A  | \lambda_j  \rangle \,, \label{eq:lambda_j_def}
\end{equation}
where the last identity follows from \eqref{eq:radial_angular}, as the ``Schr{\"o}dinger equation'' corresponding to the unitary evolution $U$. It is also useful to define the spectrum of $X$ as $X|x_a\rangle = x_a|x_a\rangle$.

We now claim that $|\lambda_i\rangle$ corresponds to a quasiparticle with momentum $\lambda_i$, which is delocalized in space as the ``wavefunction'' $|\lambda_i\rangle$, i.e., has a distribution with weight $|\langle x_a |\lambda_i\rangle|^2 $ at $x_a$. Equivalently, we identify the density $\rho_p(x)$ with the local density of states of $L$ (which we shall call the empirical quasiparticle density) in the hydrodynamic limit:
\begin{align}
 &\rho_p^{\text{emp}}(x)  \longrightarrow \rho_p(x)  \text{ in the hydrodynamic limit, where}  
 \label{eq:empdos} \\
 &    \rho_p^{\text{emp}}(x) := \sum_{a=1}^{N} \sum_{j = 1}^N 
    |\langle x_a |\lambda_j\rangle|^2 \delta(x-x_a) \delta(p-\lambda_j)  \label{eq:rho_p_emp} \,.
\end{align}
Eq. \eqref{eq:empdos} is the main result of this paper: it relates a microscopic configuration of the Calogero model to a macroscopic description in terms of the quasiparticle density. In what follows, we shall justify it with both analytical and numerical arguments. 

\subsection{Relation to thermodynamic Bethe Ansatz}
A first consequence of the Bethe-Lax correspondence \eqref{eq:rho_p_emp} is that the density of states of the Lax matrix, averaged over thermal configurations, is given by the solution of the classical Yang-Yang equation, see Eqs. \eqref{eq:YY} and \eqref{eq:rho_p_classical}:
\begin{align}
\label{eq:chargeDOS}
\widetilde{\rho}_p = \lim_{\substack{N,L\to \infty, \\N/L= \mathrm{const}}} \overline{\frac{1}{L} \sum_{j=1}^N \delta(p-\lambda_j)}  \,.
\end{align}
where the bars denote thermal averages and the particles are assumed to be confined to a large box of length $L$ (this is necessary to define a thermal state, since otherwise the particles are unbounded~\cite{PolychronakosReview}).

Eq.~\eqref{eq:chargeDOS} can be checked analytically at zero temperature, at which the particles are immobile ($\dot{x}_a = 0$) and form a lattice, $x_a - x_b = (a-b) / \rho$, where $\rho = N / L$ is the density. In the $N \to \infty$ limit, the Lax matrix is a circulant matrix and can be diagonalized by Bloch plane waves in the Brillouin zone $[-\pi \rho, \pi \rho)$. As a result, we obtain a constant density of states in the interval $[-\ell \rho \pi, \ell \rho \pi]$, in agreement with the prediction of Eq.~\eqref{eq:rho_p_classical} (with $\beta \to \infty$ and suitable chemical potential).

\begin{figure}
    \centering
    \includegraphics[width=.7\textwidth]{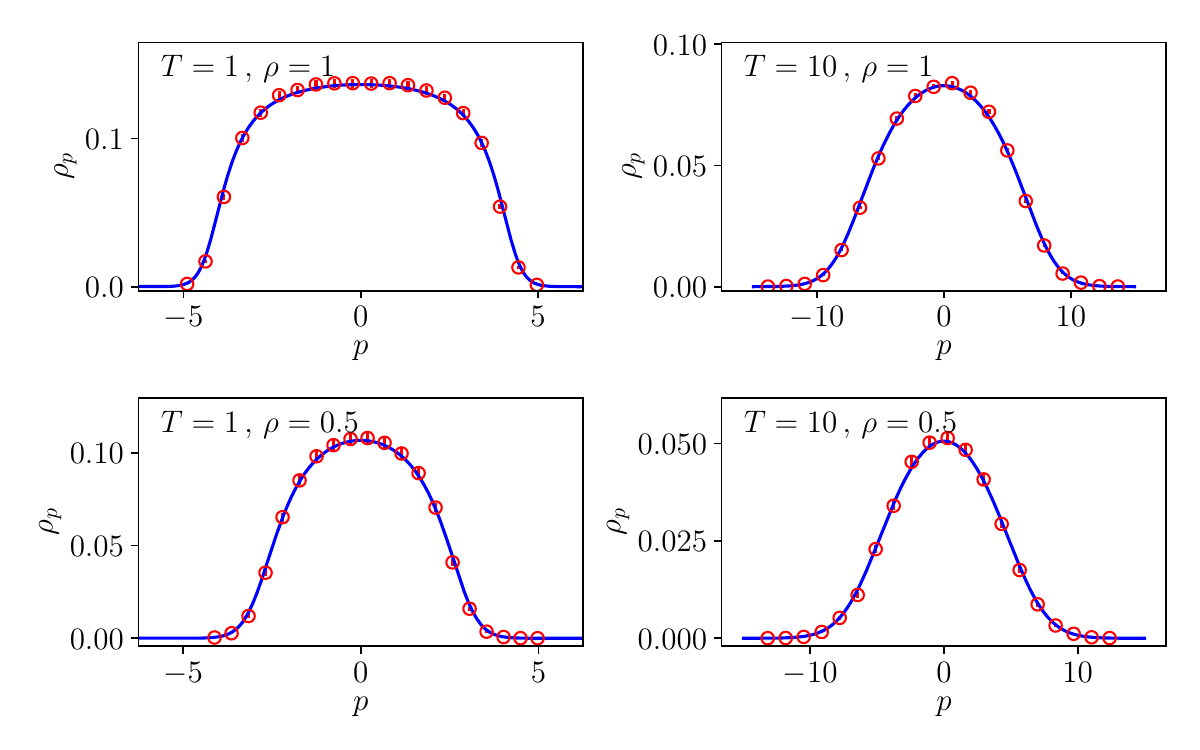}
    \caption{Thermal quasiparticle density from the Lax matrix (markers) vs predictions from thermodynamic Bethe-Ansatz (TBA), Eqs. \eqref{eq:YY} and \eqref{eq:rho_p_classical}. For the Lax matrix data, we confine $N = 256$ particles in a box of length $L = N / \rho$ ($\rho = 1$ and $1/2$) and generate $10^4$ thermal states of temperature $1/\beta = 1$ and $10$ using the standard Markov-chain Monte Carlo method. We diagonalize the resulting Lax matrices and plot the averaged density of states (normalized so that $\int dp \, \rho_p = \rho$).  To obtain the TBA prediction, we solve the Yang-Yang equation numerically and adjust the chemical potential $\mu^{cl}$ to match $\rho$. We set $\ell = 1$ for all comparisons.}
    \label{fig:thermal}
\end{figure}

At non-zero temperature, a proof of \eqref{eq:chargeDOS}, as was achieved for the particle density of states for the Toda lattice by analogy with Dumitriu-Edelman random matrix ensembles~\cite{Spohn_2019}, seems beyond reach at present. Nevertheless, we can check this equation numerically by generating thermal states from classical Monte Carlo simulations and diagonalizing the $L$ matrix. We observe good agreement with numerics at various non-zero temperatures and densities. See Fig.~\ref{fig:thermal} for detailed methods and a sample of results. 

\subsection{Boltzmann equation}\label{sec:boltzman}
A further important check of the Bethe-Lax correspondence is to show that the empirical quasiparticle density $ \rho_p^{\text{emp}}(x)$ \eqref{eq:rho_p_emp} satisfies the Boltzmann equation \eqref{eq:minmod}, subject to an appropriate approximation in the hydrodynamic limit. 

To this end, we rewrite the empirical quasiparticle density as \begin{equation}
    \rho_p^{\text{emp}}(x) = \mathrm{Tr}\left[ \delta(p - L) \delta(x - X)  \right] \,,\, \text{where } \delta(p - L) = \frac1{2\pi i} \left[ (p - i \epsilon - L)^{-1} - (p + i \epsilon - L)^{-1}  \right]_{\epsilon\to 0_+} \,, \label{eq:Stie}
\end{equation}
and similarly for $\delta(x - X)$, by the Stieltjes inversion formula. Expanding the matrices inverses at complex infinity, the above formula can be viewed as a resummation of the moments $\mathrm{Tr}[X^n L^m]$. To calculate the time derivative, it is simplest to work with unrotated operators Eq. \eqref{eq:matrixEOM} and rotate back after differentiation. This yields
\begin{align}
    \nonumber &\partial_t \mathrm{Tr}[(p-L)^{-1}  (x-X)^{-1}] = \mathrm{Tr}[(p-L)^{-1} F(X) (p-L)^{-1}  (x-X)^{-1}] + 
     \mathrm{Tr}[(p-L)^{-1} (x-X)^{-1} L (x-X)^{-1}]  \\
     \nonumber =& \mathrm{Tr}[(p-L)^{-2}  F(X) (x-X)^{-1}] + 
     \mathrm{Tr}[(p-L)^{-1}  L (x-X)^{-2}] + \mathcal{O}([X, L]) \\
     = & - \partial_p \mathrm{Tr}[(p-L)^{-1}  F(X) (x-X)^{-1}] - 
   \partial_x \mathrm{Tr}[(p-L)^{-1}  L (x-X)^{-1}] + \mathcal{O}([X, L]) 
\end{align} 
In the second line, we invoked a semiclassical approximation consisting of ignoring the commutator $[X,L]$ (and, assuming $F$ is analytical, $[F(X), L]$). We shall discuss the physical interpretation and validity of this step below. Assuming its validity for now and substituting into Eq.~\eqref{eq:Stie}, we have
\begin{align}
 \nonumber \partial_t \rho_p^{\text{emp}}(x) &= - \partial_p \mathrm{Tr}[\delta(p-L)  F(X) \delta(x-X)] - 
   \partial_x \mathrm{Tr}[\delta(p-L) L \delta(x-X)] + \mathcal{O}([X, L]) \\
   \label{eq:empEOM}
   & = - (p\partial_x + F(x) \partial_p )  \rho_p^{\text{emp}}(x)  + \mathcal{O}([X, L])  \,.
\end{align}
We thus conclude that the empirical density satisfies the Boltzmann equation with the na{\"i}ve forcing term, assuming the semiclassical approximation.

We now discuss the meaning of this approximation, and provide some arguments in its favour. First, we focus on the case without a trap, $F = 0$ (the trapped case will be discussed in Section~\ref{sec:trap} below). With no trap, we can show that this approximation captures the exact time evolution of the first and second moments in position of the empirical quasiparticle density. More concretely, the moments are defined as 
\begin{equation}
   \left<x^n\right>_p := \int dx\, x^n \rho_p^{\text{emp}}(x) \,.
\end{equation} 
If $ \rho_p^{\text{emp}}(x)$ satisfies the non-interacting Boltzmann equation, we would expect $\frac{d}{dt} \left<x^n\right>_p = p n \left< x^{n-1} \right>_p$. Starting from the definition \eqref{eq:rho_p_emp} and using the equations of motion Eqs.~\eqref{eq:redEOM1} and \eqref{eq:redEOM2}, it is not hard to check that this is indeed the case for $n \le 2$:
\begin{equation}
   \frac{d}{dt}  \left<x^n\right>_p  = p n \left< x^{n-1} \right> \,,\, n \le 2 \,,\, F = 0 \,.
\end{equation}
In particular, $\partial_t \left< x\right>_p = \rho_p^J$ is the current density, so the above identity for $n = 1$ implies the current equation of state \eqref{eq:Jandp} for homogeneous states. Usual arguments then imply that the forceless Boltzmann equation Eq.~\eqref{eq:class_boltzmann} is satisfied if the empirical quasiparticle density varies slowly in space and time: the error term $\mathcal{O}([X, L])$ corresponds to subballistic corrections. To see this more explicitly, consider the $n = 3$ moment (similar arguments apply for $n > 3$, though their complication increases with $n$):
\begin{align}
    \partial_t \left<x^3 \right>_p &= 3 p \mu_{2} + D_2 \,,\, \label{eq:boltzmann_diffu} \\
    D_2 &=  \mathrm{Tr}[\delta(p - L) [X, L] X ] \\
    \label{eq:boltzmann_diffu3}
    &=  
   i\ell\sum_{j=1}^N \delta(p - \lambda_j)  \langle \lambda_j \vert u\rangle \langle u \vert X \vert \lambda_j \rangle  - i \ell \left< x \right>_p
\end{align}
where $|u\rangle = (1,1, \dots, 1)^T$.  On the right hand side of \eqref{eq:boltzmann_diffu}, the ballistic term $\mu_{n-1} = \mathrm{Tr}[\delta(p-L) X^{n-1}]$, while the correction $D_n$ involve $(n-2)$ powers of $X$. Now, in a system with $N \gg 1$ particles at order $1$ density, $\Vert X \Vert \sim N$ as an operator; hence prima facie, $D_n$ is of order $1/N$ compared to the ballistic term. There is however a caveat: the vector $\vert u \rangle$ has norm $\sqrt{N}$. Nevertheless, its overlap with quasiparticle wavefunctions $|\langle u \vert  \lambda_j \rangle|^2$ is large ($\sim  N$ at most) only if the wavefunction is delocalized; for localized wavefunctions, the overlap is $1/N$ smaller, of order unity. In the latter case, the term involving $\vert u\rangle$ is also subleading in the hydrodynamic limit.

Thus the semiclassical approximation relies on the localization of quasiparticle eigenstates. (This is after all expected, since by basic quantum mechanics $[X,L]$ measures the position uncertainty of the quasiparticle eigenstates.) Now, at nonzero temperature, the thermal ensemble of Lax matrices $L$ closely resembles a power-law random matrix ensemble studied recently~\cite{localization}, for which it is shown that most eigenstates are localized algebraically: $|\langle x_a|\lambda_j\rangle|^2 \sim |x_a - x|^{-\gamma}$ for $ \gamma \approx 2$, where $x$ is the localization centre of the wavefunction. On the other hand, as $T \to 0$, the randomness disappears and the quasiparticles delocalize. We therefore expect the most stringent dynamical tests of the Bethe-Lax correspondence to be quenches from zero/low temperature states, in presence of spatial inhomogeneity. 

We now offer some more speculative remarks on the validity of the non-interacting Boltzmann description Eq. \eqref{eq:empEOM}. First, notice that the commutator correction to the $n>2$ moment equations, as in Eq. \eqref{eq:boltzmann_diffu3}, while subleading in $N$ also scales with the interaction strength $\ell$. Thus at the level of hydrodynamics, the Calogero interaction strength controls the strength of higher-order corrections to the non-interacting Boltzmann description. In the standard Calogero model, the interaction strength $\ell$ is taken to be independent of $N$. However, Eq. \eqref{eq:boltzmann_diffu3} suggests a possibility of accessing different dynamical scaling regimes by varying the scaling of $\ell$ with $N$.

We next note that although the empirical distribution function does not, in general, satisfy the non-interacting Boltzmann equation exactly, there is another distribution function,
\begin{align}
\label{eq:exactDOS}
\rho^{\text{ex}}_{p}(x) = \sum_{j=1}^N \delta(x - \langle X \rangle_j) \delta(p-\lambda_j)
\end{align}
where $\langle X\rangle_j = \langle \lambda_j | X | \lambda_j \rangle$,
which \emph{does} satisfy Eq. \eqref{eq:class_boltzmann} exactly when $F=0$, as a consequence of the microscopic equations of motion Eqs. \eqref{eq:redEOM1} and \eqref{eq:redEOM2}. On the other hand, Eq. \eqref{eq:exactDOS} is somewhat unphysical, in the sense that it is non-local in the bare particle coordinates $x_a$. Nevertheless, a comparison with the empirical density of states Eq. \eqref{eq:empdos}, together with the identity $|\langle x_a | \lambda_j \rangle|^2 = \partial \langle X \rangle_j/\partial x_a$, suggests the identification
\begin{equation}
\label{eq:conj}
\delta(x- \langle X \rangle_j) \approx \sum_{a=1}^N  \frac{\partial \langle X \rangle_j}{\partial x_a} \delta(x-x_a) \,.
\end{equation}
This conjectural formula relates the local density of a single quasiparticle to the local densities of all bare particles. Its Jacobian form is reminiscent of the idea that quasiparticle dressing in integrable systems is equivalent to a nonlinear coordinate transformation from free particles~\cite{Percus69,DoyonGeometry,DoyonSoliton,Bulchandani_2017,cardy2021toverline}. By the previous discussion, we expect that the error in Eq. \eqref{eq:conj} is also controlled by the degree of quasiparticle localization, i.e. the magnitude of the commutator $[X,L]$. However, since Eq. \eqref{eq:exactDOS} does not generalize readily beyond harmonic trapping potentials (unlike the more approximate, but more physical, empirical quasiparticle density), we shall not pursue its investigation further in this paper.

\subsection{Two-reservoir initial condition}

\begin{figure}
    \centering
    \includegraphics[width=.85\textwidth]{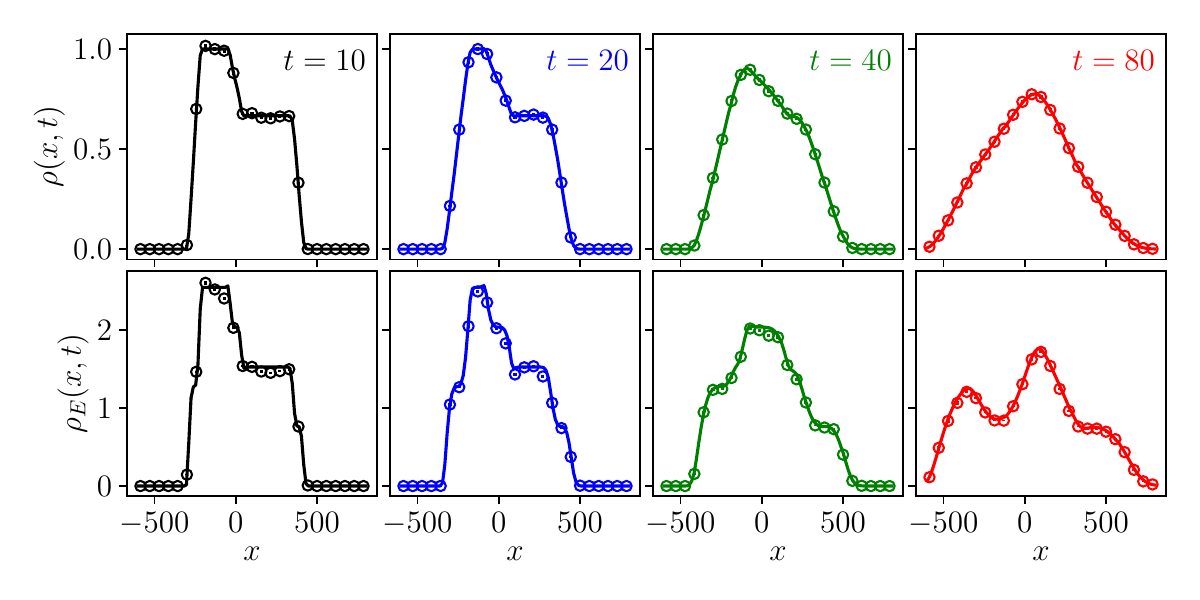}
    \caption{Time evolution of particle and energy density starting from a two-reservoir initial condition. Left reservoir: $N = 256$ particles confined to $[-N, 0]$ at $T = 1$. Right reservoir:  $256$ particles confined to $[0, 3N/2 ]$ at $T = 2$. The numerical data (circles) is averaged over $400$ thermal samples. The analytical prediction (solid curves) is obtained from the TBA equation and the free Boltzmann equation, see Eq. \eqref{eq:twores_pred}. We set $\ell = 1$ for all plots.}
    \label{fig:two_res}
\end{figure}

To conclude this section, we test the kinetic theory of $\rho^{\mathrm{emp}}_p(x)$ numerically in a standard ``two-reservoir'' setting. For this purpose, we prepare two different equilibria confined to the boxes $[-A, 0]$ and $[0, B]$ respectively (usually the reservoirs are considered semi-infinite; yet such a limit is hard to attain with the Calogero long-range interaction)    . At $t = 0$, the particles are released. We then measure the time evolution of particle and energy density under the Calogero dynamics (with $F = 0$). We note that the long-range Calogero interaction means that an \textit{a priori} definition of the energy density is not obvious. We define it using the Bethe-Lax correspondence, as 
\begin{equation}
     \rho_E(x) = \int dp \, \frac{1}{2}p^2 \rho^{\text{emp}}_p(x)
\end{equation}
The numerical data are then compared with the non-interacting kinetic theory prediction
\begin{equation}
    \rho(x,t) = \int_{x/t}^{(x+A)/t} dp \, \rho_L(p)  + \int_{(x-B)/t}^{x/t} dp \, \rho_R(p) \label{eq:twores_pred}
\end{equation}
where $\rho_L$ and $\rho_R$ are the quasiparticle density of the reservoirs, obtained by TBA. An excellent agreement is found, see Fig.~\ref{fig:two_res}. This is not trivial given that the temperatures of the reservoirs are low enough that the localization length of the quasiparticle wavefunctions (at $t = 0$) is about half the box size. Even so, the semiclassical approximation appears to be robust. 

\section{Trapped dynamics}\label{sec:trap}
We argued above that the quasiparticle density of the Calogero model, defined via the Bethe-Lax correspondence, satisfies a non-interacting Boltzmann equation even in the presence of trapping potentials. In this section, we test the non-interacting Boltzmann description against microscopic simulations of dynamics in harmonic and anharmonic traps. We find excellent agreement in all cases, even for complicated nonlinear phenomena such as soliton dynamics in the harmonically trapped Calogero-Moser model\cite{OldCalog3}.

\subsection{Harmonic trap}
As recalled in Section \ref{sec:classical}, the classical Calogero model remains integrable in the presence of a harmonic trapping potential, $V(x) = \frac{1}{2} \omega^2 x^2$. In the matrix model formulation Eq. \eqref{eq:matrixEOM}, the corresponding potential reads $V(M) = \frac{1}{2} \omega^2 M^2$. Thus $F(M) = -\omega^2 M$ and the variable
\begin{equation}
     Q = \omega M - i  \dot{M},
\end{equation} 
simply rotates: 
\begin{equation}
    \dot{Q} = i \omega Q \Rightarrow Q(t) = e^{i \omega t} Q(0) \,. \label{eq:Q}
\end{equation}
From periodicity of $Q$, it is immediate that the dynamics of the harmonically trapped Calogero model exhibits ``perfect revivals'' at each trap period, $T = 2\pi/\omega$.

Intriguingly, this property is shared by the na{\"i}ve kinetic theory Eq. \eqref{eq:minmod}, which for a harmonic trap takes the form
\begin{equation}
\label{eq:harmonictrap}
\partial_t \widetilde{\rho}_p + p \partial_x \widetilde{\rho}_p - \omega x \partial_p \widetilde{\rho}_p = 0,
\end{equation}
and has an exact solution by characteristics, with characteristic curves given by
\begin{align}
\label{eq:char}
    \frac{d}{dt} x = p, \quad \frac{d}{dt} p = -\omega x,
\end{align}
corresponding to uniform rotation of the distribution function $\rho$ with frequency $\omega$ on constant-energy ellipses, $p^2 + \omega^2 x^2 = 2E$.

The Bethe-Lax correspondence provides a direct relation between the Boltzmann equation Eq. \eqref{eq:harmonictrap} and the microscopic motion. Indeed, it predicts that the empirical quasiparticle density $\rho^{\text{emp}}$ rotates in the simple manner described above. In Section~\ref{sec:boltzman}, we showed that $\rho^{\text{emp}}$ satisfies Eq \eqref{eq:harmonictrap} under the semiclassical approximation but only discussed its justification in the case $F = 0$. For a harmonic trap $F = \omega X$, the situation is similar. The na{\"i}ve kinetic theory \eqref{eq:harmonictrap} again yields the exact time evolution of the lowest moments, but one must now consider moments in both $x$ and $p$. Indeed, defining
\begin{equation}
    \left< x^n p^m \right> := \int dx \, dp \, x^n p^m \rho^{\text{emp}}_p(x) = \mathrm{Tr}[X^n L^m] \,,\,
\end{equation}
the equations of motion \eqref{eq:redEOM1}, \eqref{eq:redEOM2} imply that
\begin{equation}
   \frac{d}{dt}\left< x^n p^m \right> = n \left< x^{n-1} p^{m+1} \right> - m \omega \left< x^{n+1} p^{m-1} \right>   \,,\, n \le 2, m \le 2 \,, \label{eq:moment_trap}
\end{equation} 
which can also be derived by assuming  $\rho^{\text{emp}}_p(x)$ satisfies \eqref{eq:harmonictrap}. For higher moments, the above equation still holds modulo a correction which has a commutator $[X, L]$ and $n + m - 1$ powers of $X$ and $L$, which is one less compared to the RHS of \eqref{eq:moment_trap}. The argument then proceeds as in the flat case, provided one observes that in presence of the trap, both $X$ and $L$ have eigenvalues of order $\sqrt{N}$ typically (recall that in contrast, $L \sim \mathcal{O}(1)$ and $X \sim \mathcal{O}(N)$ typically in the flat case). Hence we expect finite-size effects to be more pronounced in the presence of a trapping potential.

\begin{figure}
    \centering
    \includegraphics[width=.7\columnwidth]{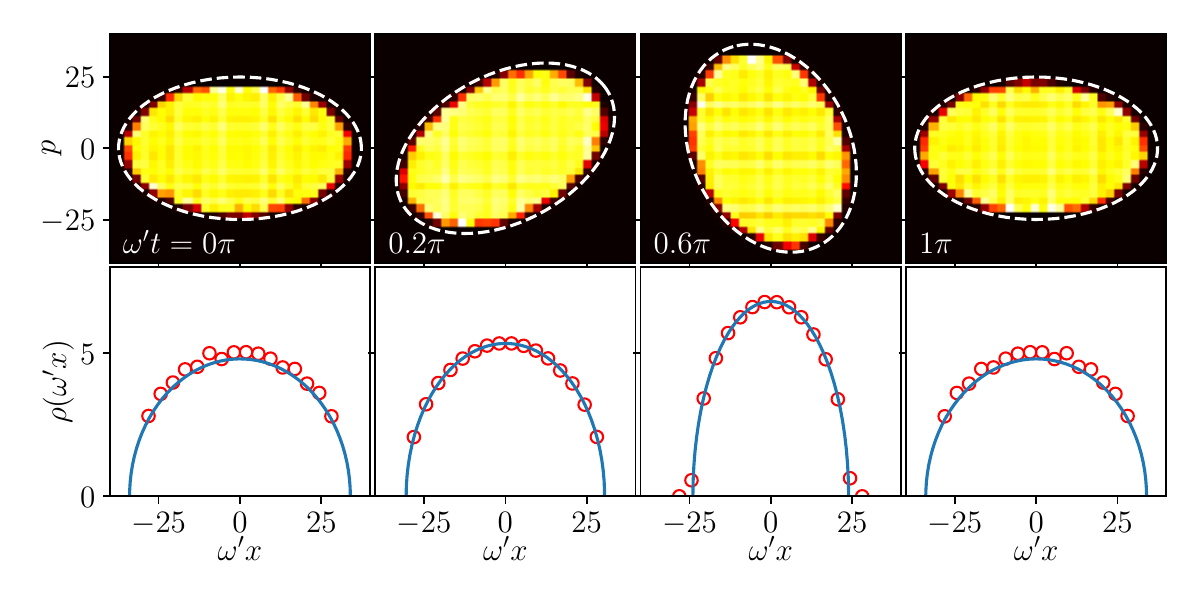}
    \caption{Top panels: the empirical quasiparticle density in a harmonic trap with frequency $\omega' = 1.5$, evolving from an initial condition consisting of a thermal cloud of $256$ particles at temperature $T = 1$ and confined to a harmonic trap with frequency $\omega = 1$. The dashed curves show an ellipse that rotates with angular velocity $\omega'$. The bottom panels compare the measured particle density (circles, average of $200$ thermal samples) to the na{\"i}ve kinetic theory prediction (solid curves). We set $\ell = 1$ for all plots.}
    \label{fig:trap_thermal}
\end{figure}

We now test and illustrate the Bethe-Lax correspondence with two numerical examples. First, we consider a quench in which a Calogero-Moser gas is prepared at low temperature $T$ in a trap of frequency $\omega$, and at $t = 0$, the trap frequency is quenched to $\omega' \ne \omega$. We then compute the empirical quasiparticle density at different times (by solving the dynamics using Eq.~\eqref{eq:Q} and diagonalizing the Lax matrix). Plotting this in the $(\omega' x,p)$ phase space, we can clearly see that the empirical quasiparticle density rotates simply as predicted by the na{\"i}ve kinetic theory, see Fig.~\ref{fig:trap_thermal}. We also verify that this picture yields quantitatively correct predictions for the local particle density, despite the pronounced finite-size effect (in particular near the edge of the distribution). 

As a more spectacular example, we consider a two-soliton solution, revealed in Ref.~\citen{OldCalog3}. The initial conditions for the two soliton solution are chosen as follows. We choose two dual complex variables $z_1$ and $z_2$. The initial conditions are chosen such that the following conditions are satisfied:
\begin{eqnarray}
\label{eq:two_sol_x}
\omega x_j &=& \ell \sum_{k=1 (k \neq j)}^N \frac{1}{x_j - x_k}-\frac{\ell}{2}\bigg(\frac{1}{x_j-z_1} + \frac{1}{x_j-\bar{z}_1} +\frac{1}{x_j-z_2} +\frac{1}{x_j-\bar{z}_2} \bigg) \\
p_j &=&\frac{i\ell}{2} \bigg(\frac{1}{x_j-z_1} - \frac{1}{x_j-\bar{z}_1} +\frac{1}{x_j-z_2} -\frac{1}{x_j-\bar{z}_2}  \bigg)
\label{eq:two_sol_p}
\end{eqnarray}

\begin{figure}
    \centering
    \includegraphics[width=.55\textwidth]{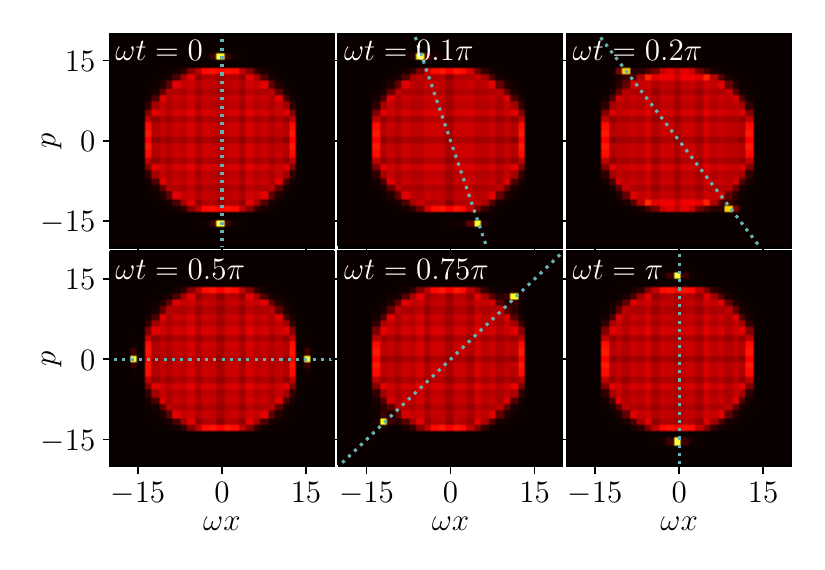}
    \includegraphics[width=.3\textwidth]{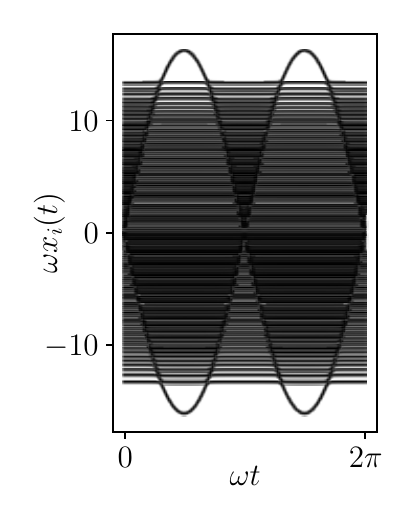}
    \caption{Left panels: Time evolution of the empirical quasiparticle density of a two-soliton solution \eqref{eq:two_sol_x}, \eqref{eq:two_sol_p}, with $\ell = 1$, $\omega = 1$, $z_1 = -z_2 = 0.251247i$, $N = 101$. The dashed line rotates with angular velocity $\omega$. The solitons appear as peaks in the $(x, p)$ plane, rotating with angular velocity $\omega$. Right panel: the evolution of particle positions of the same solution, over a period. Note that the trajectories do not cross each other; the solitons are a collective phenomenon involving all particles. }
    \label{fig:twosolition}
\end{figure}

For fixed $z_1$ and $z_2$, Eq.~\eqref{eq:two_sol_x} and Eq.~\eqref{eq:two_sol_p} are non-trivial to solve. However, one can employ a damping equation~\cite{kulkarni2017emergence,gon2019duality} that leads to a potential minimization problem. Once we find the initial condition for the two-soliton case, the subsequent evolution can be performed via the matrix formulation given in Eq.~\eqref{eq:Q}. We then numerically compute the empirical quasiparticle density of the two-soliton solution and plot it in Fig.~\ref{fig:twosolition}. Remarkably, the solitons manifest themselves as sharp peaks on either side of the disk distribution (which corresponds to the zero-temperature state in a harmonic trap). The peaks appear near the edge of the disk and rotate in phase space as if they were free particles in the harmonic potential, never encountering one another.

\subsection{Nonintegrable anharmonic traps}
A distinctive feature of the conjectured Bethe-Lax correspondence is that it can be formulated for general anharmonic traps, even non-integrable ones: the empirical quasiparticle density, defined in Eq.~\eqref{eq:rho_p_emp}, should satisfy the non-interacting Boltzmann equation Eq. \eqref{eq:minmod}. By the arguments in Section~\ref{sec:boltzman}, this statement is true under the ``semiclassical approximation'' in which commutators of the form $\mathcal{O}([X,L])$ are neglected. However, the dynamics of $X$ and $L$ depend on the shape of the potential under consideration, in such a way that it is difficult to formulate general analytical arguments beyond the harmonic case.

We therefore proceed directly to a numerical test. To this end, we prepare a zero-$T$ cloud in a harmonic potential $V(x) = \frac{1}{2} x^2$ and then quench to an anharmonic potential, $V(x) = 2 \sqrt{1+x^2}$, that is not expected to preserve integrability. We see that the non-interacting Boltzmann equation Eq. \eqref{eq:minmod} successfully captures the relaxation dynamics, which therefore consists solely of single-(quasi)particle dephasing on accessible time-scales.
\begin{figure}
    \centering
    \includegraphics[width=.6\columnwidth]{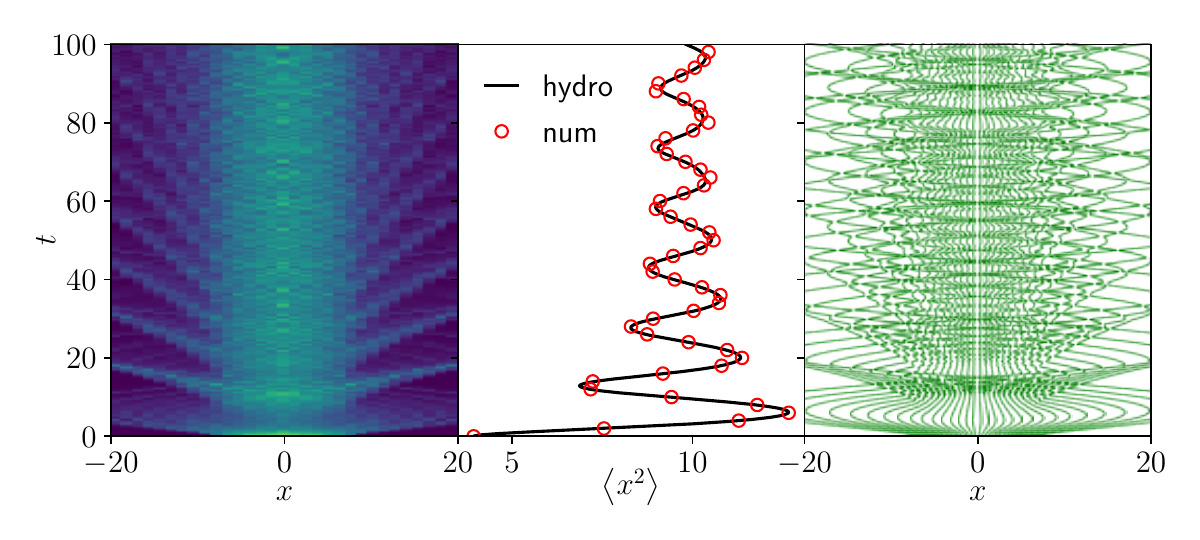}
    \caption{Left: non-interacting Boltzmann prediction for the particle density. Middle: comparison between hydrodynamic and microscopic predictions for the mean squared width of the packet. Right: microscopic evolution. Model parameters are set to $N = 32$, $\ell = 1$. }
    \label{fig:aharm}
\end{figure}

\begin{figure*}
    \centering
    \includegraphics[width=.6\textwidth]{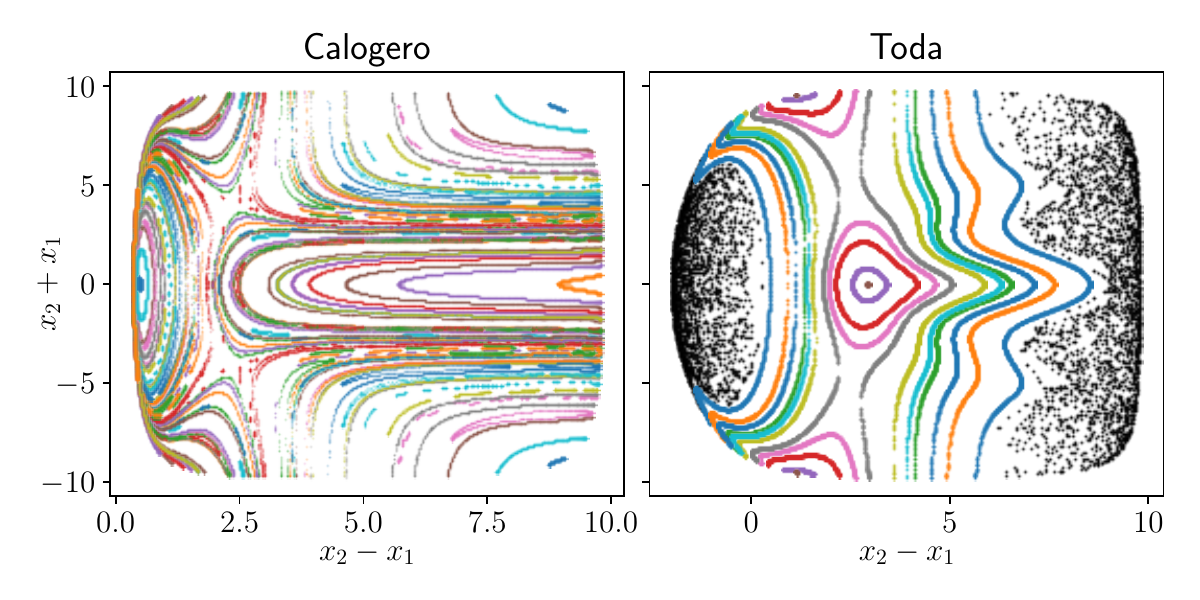}\\
    \includegraphics[width=.6\textwidth]{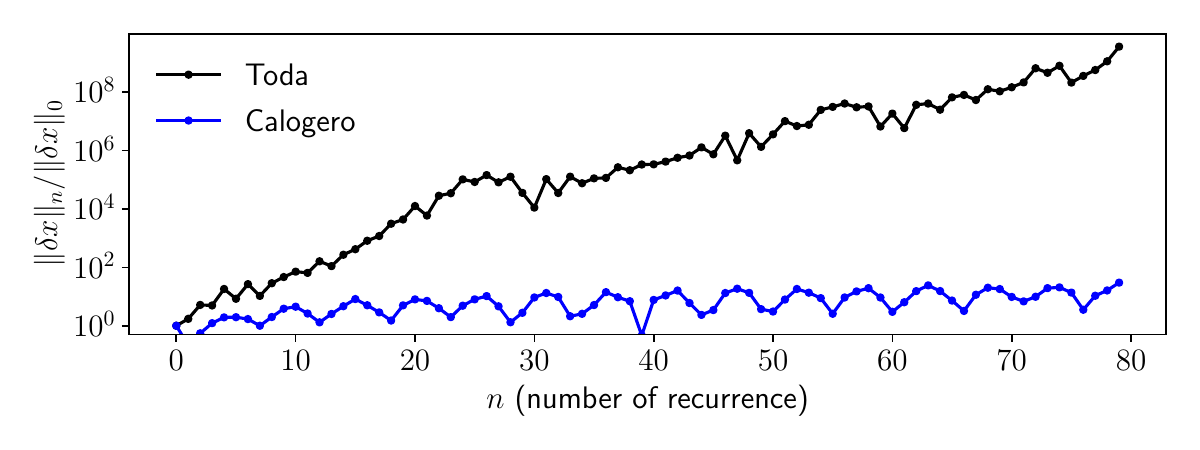}
    \caption{Top panels: Orbits of the recurrence map in the Poincar\'e section defined by $E = 10$ and $p_1 = -p_2$, in a system with two particles in an external potential $V(x) = \sqrt{x^2+1}$ and interacting with potentials $1/(x_1-x_2)^2$ (Calogero, left) and $e^{-(x_1-x_2)}$ (Toda, right). Different colors are assigned randomly to trajectories to distinguish them. Integrability of the Toda interaction is broken by the trap: the black points represent a \textit{single} chaotic trajectory that is ergodic in a large portion of the phase space. Meanwhile, that of the Calogero interaction seems robust.
    Bottom panel: Growth of the separation $\Vert \delta x \Vert_n = |x_1(n) - x'_1(n)| + |x_2(n) - x'_2(n)|$ between a pair of nearby orbits ($x_i$ and $x'_i$, separated by an initial distance $\Vert \delta x \Vert_0 = 10^{-8}$) under either dynamics. In the Toda case, the orbits start from $x_1(0) + x_2(0) = 0.1, x_1(0) - x_2(0) = -0.5$ in the chaotic basin, and exhibit exponential growth with a positive Lyapuov exponent. In the Calogero case, we found no such chaotic trajectory after an exhaustive search. The illustrated orbit starts from $x_1(0) + x_2(0) = 0.1, x_1(0) - x_2(0) = 3$. 
    }
    \label{fig:poincare}
\end{figure*}

The performance of the simple kinetic theory in an integrability-breaking trap is intriguing. On general grounds, we expect the agreement to break down eventually as the dynamics becomes chaotic~\cite{CBMTrap}.  However, for the trapped Calogero model we were unable to observe the onset of chaos on accessible time-scales. This is probably due to the robustness of the Calogero model's integrability to fourth-order trapping potentials (cf. the discussion around Eq. \eqref{eq:arbtrap}), which suggests a robustness to trapping potentials in general, compared to other integrable models. This expectation is borne out by numerical simulations of Poincar{\'e} sections for the two-particle Calogero model in an integrability-breaking trap compared to the two-particle Toda chain in such a trap, see Fig. \ref{fig:poincare}.
It is also consistent with studies of the Toda chain in power-law pinning potentials~\cite{PinToda1,PinToda2}, which exhibit a crossover to normal diffusion at large system sizes, at variance with our observations of purely ballistic evolution for the Calogero model.

\section{Discussion}

We have derived the kinetic theory of Calogero quasiparticles on a line, which reduces to the non-interacting Boltzmann equation. We showed that this simplification could be understood from an emergent quasiparticle description in terms of eigenvectors of the Lax matrix. The resulting expression for the empirical single-particle distribution function, Eq. \eqref{eq:empdos}, yields excellent agreement with the non-interacting Boltzmann equation on accessible time-scales, even in the presence of integrability-breaking trapping potentials.

One point which merits further comment is the long-ranged character of interactions in the Calogero model. At first sight, such long-ranged interactions render the conserved charges and currents of the Calogero model non-local~\cite{SpohnIntNonInt}, which calls into question the validity of a local, hydrodynamic description. At the same time, the Calogero model is exactly solvable using the asymptotic Bethe ansatz\cite{SutherlandAsymp}, suggesting that the scattering behaviour of its quasiparticles is much the same as for short-range-interacting integrable models. Indeed, the validity of hydrodynamics for the Calogero model is most easily justified in the quasiparticle language: according to the Bethe-Lax correspondence presented above, dynamics in the Calogero model is captured by quasiparticles that are localized in space at non-zero temperature, despite the long-ranged interactions between bare particles. In this sense, the quasiparticle kinetic theory of the Calogero model is no different from that of a short-ranged integrable model.

We noted earlier that there is a mature theory of the zero-temperature (``superfluid'') hydrodynamics of quantum and classical Calogero models based on collective field theory~\cite{OldCalog1,OldCalog2,stone2008classical,OldCalog3}. When subballistic derivative corrections to the latter are neglected, it matches the zero-temperature limit of our results. We anticipate that a systematic treatment of the commutator corrections to the free Boltzmann evolution in Eq. \eqref{eq:empEOM} (or perhaps a first-principles treatment of Eq. \eqref{eq:conj}) would recover these subballistic derivative corrections at non-zero temperature. Another interesting common feature between our analysis and the collective field theory of Calogero models is that the latter remains integrable in arbitrary external trapping potentials\cite{POLYCHRONAKOS1992341}, just like the non-interacting Boltzmann equation Eq. \eqref{eq:minmod}. Finally, a natural extension of our work would be to develop a kinetic theory for the broader family of compactified Calogero-Sutherland models with trigonometric or hyperbolic interactions, which inherit the property of remaining integrable in suitably chosen trapping potentials\cite{kulkarni2017emergence}. It seems plausible that a Bethe-Lax correspondence analogous to Eqs. \eqref{eq:rho_p_emp} and \eqref{eq:chargeDOS} holds for all of these models.

Our results demonstrate that the integrability-breaking effects of external trapping potentials, as arise in present-day experiments on ultracold Bose gases~\cite{cradle,Dubail,malvania2020generalized}, can be rather subtle. For example, the absence of diffusion on the Navier-Stokes scale for both the quantum and classical Calogero models seems to imply that the mechanism of diffusive thermalization proposed recently for trapped integrable gases~\cite{BastianelloTrap,durnin2021diffusive} does not apply. A related question is whether the trapped Calogero model exhibits a finite ``time to chaos'', that was observed for systems of trapped classical hard rods~\cite{CBMTrap} but appears to be numerically inaccessible for the Calogero model, even in traps that are expected to break integrability. Both findings suggest that the tendency to chaos for trapped Calogero particles, quantum or classical, is remarkably weak. Thus identifying the signatures of broken integrability in the Calogero model might pave the way towards a deeper understanding of broken integrability in the quantum Newton's cradle~\cite{cradle}.\\

\textit{Acknowledgments.} We are grateful to A.G. Abanov for helpful discussions and to H. Spohn for collaborations on related topics. We thank S. Santra for help with data for thermal initial conditions for the trapped case. MK would like to acknowledge support from the project 6004-1 of the Indo-French Centre for the Promotion of Advanced Research (IFCPAR), Ramanujan Fellowship (SB/S2/RJN-114/2016), SERB Early Career Research Award (ECR/2018/002085) and SERB Matrics Grant (MTR/2019/001101) from the Science and Engineering Research Board (SERB), Department of Science and Technology, Government of India. MK acknowledges support of the Department of Atomic Energy, Government of India, under Project No. RTI4001.  J.E.M. acknowledges support from US National Science Foundation grant DMR-1918065.

\bibliography{calogbib}
\end{document}